\documentstyle[prl,aps,multicol,epsf]{revtex}
\begin{document}
\draft
\widetext
\title{Ferromagnetic transition in a double-exchange model}
\author{Eugene Kogan$^{1}$, Mark Auslender$^2$ and Eran Dgani$^{1}$}
\address{$^1$ Jack and Pearl Resnick Institute 
of Advanced Technology,
Department of Physics, Bar-Ilan University, Ramat-Gan 52900, 
Israel\\
$^2$ Department of Electrical and Computer Engineering,
Ben-Gurion University of the Negev,
P.O.B. 653, Beer-Sheva, 84105 Israel}
\date{\today}
\maketitle
\begin{abstract}
\leftskip 54.8pt
\rightskip 54.8pt
We calculate the temperature  of a ferromagnetic transition in a
double-exchange model with classical core spins for arbitrary relation
between Hund exchange coupling and electron band width by solving
the Dynamical Mean 
Field Approximation equations. 
\end{abstract}
\begin{multicols}{2}
\narrowtext

\section{Introduction}

The  double-exchange (DE) 
model \cite{zener,anderson,degennes} is one of the basic ones 
in the theory of magnetism. Magnetic ordering appears in this model  due to  
Hund 
exchange coupling  between the core
spins and the mobile carriers. The Hamiltonian  of the model is 
\begin{eqnarray}
\label{HamDXM}  
H = \sum_{nn'\alpha} t_{n-n'} c_{n\alpha}^{\dagger} c_{n'\alpha}
-J \sum_{n\alpha\beta} {\bf S}_n\cdot 
{\bf \sigma}_{\alpha\beta}c_{n\alpha}^{\dagger} c_{n\beta},
\end{eqnarray}
where $c$ and $c^{\dagger}$ are the electrons annihilation and creation
operators, ${\bf S}_n$ is the operator of core spin,  $t_{n-n'}$ is the 
electron hopping,  
$J$ is  Hund 
exchange 
coupling between a core spin and a conduction electron,
$\hat{\bf \sigma}$ is the vector of the Pauli matrices, and $\alpha,\beta$ are
spin indices.

The model (the core spins being treated as classical
vectors) was thoroughly studied already in the 
papers \cite{zener,anderson,degennes}.   During the last years, because of a 
general
interest in manganites, the model was brought in the
focus of attention, and a
lot more was achieved (see reviews \cite{coey99,izyumov01,ziese02} 
and references therein). 
However, 
some basic properties of the model are still known only partially.
For example, most of
the papers dealing with the DE model, starting from classical paper by De
Gennes \cite{degennes}, considered the DE Hamiltonian
with infinite exchange (and  with the addition of the antiferromagnetic
superexchange, which is crucial for the explanation of magnetic properties of
manganites).  

The other  extremity which was studied is the particular case
of weak exchange  (much less than the electron
bandwidth), when the DE Hamiltonian
can be reduced to 
Ruderman-Kittel-Kasuya-Yosida (RKKY) Hamiltonian (see review \cite{vf} and
references therein).

In this paper we calculate the temperature  of a ferromagnet - paramagnet 
transition $T_c$ in a
double-exchange model  for arbitrary relation
between Hund exchange coupling and electron band width by solving
the Dynamical Mean 
Field Approximation equations. 
Note that we treat the
core spins as classical
vectors. (When the quantum nature of the core spins is taken into account, 
the Hamiltonian (\ref{HamDXM}), 
which in this case is often called the periodic Kondo model, becomes much
more complicated;  only scanty results were obtained for the model up to now.)
The problem of classical spins, as we shall see, 
combines tractability with rich and
interesting physics.

\section{Hamiltonian and DMFA equations}

Like it was said above, we consider spins as classical
vectors $ {\bf S}_n = {\bf m}_n$ 
with the normalization $|{\bf m}|^2 = 1$,  Thus the DE Hamiltonian 
 in a single electron representation 
 can be presented as
\begin{equation}
\label{generic}
H_{nn'}=t_{n-n'}-J {\bf m}_n\cdot {\bf \sigma}\delta_{nn'}.
\end{equation}
We have a problem of electron scattered by core spins, 
the probability  of any given core spin 
configuration  depending upon the energy of electron
subsystem. To solve the problem we will use the
Dynamical Mean Field Approximation (DMFA) (see \cite{DMFA,furukawa2} 
and references
therein).

In this approach, first we calculate an averaged,  
with respect to random
orientation of core  spins, density of states of
electron in a random core spins configuration, treating electron scattering in a
single site approximation, and considering the probability of
any configuration as given.  We introduce Green's function
\begin{eqnarray}
\label{green}
\hat{G}(E)=(E-H)^{-1}, 
\end{eqnarray}
In this approximation the averaged   locator 
\begin{eqnarray}
\hat{G}_{\rm loc}(E)= 
\left\langle\hat{G}_{nn}(E)\right\rangle,
\end{eqnarray}
is expressed through the  the local self-energy $\hat{\Sigma}$ by the
equation
\begin{eqnarray}
\label{local}
\hat{G}_{\rm loc}(E) =g_0\left(E - \hat{\Sigma}(E)\right),
\end{eqnarray}
where
\begin{eqnarray}
\label{g}
g_0(E) =\frac{1}{N}\sum_{\bf k}\left(E-t_{\bf k}\right)^{-1} 
\end{eqnarray}
is the bare (in the
absence of the  exchange interaction) locator. The
self-energy satisfies equation
\begin{eqnarray}
\hat{G}_{\rm loc}(E)=\left\langle \frac{1}
{\hat{G}_{\rm loc}^{-1}(E)+\hat{\Sigma} (E)
+J{\bf m}\cdot\hat{\bf \sigma}}\right\rangle,
\label{cpa}
\end{eqnarray}
where $\left\langle X({\bf m})\right\rangle \equiv \int X({\bf m})P({\bf m})$,
and $P({\bf m})$ is a probability 
of a given spin orientation (one-site probability). The quantities 
$\hat{G}$ and $\hat{\Sigma}$
 are $2\times 2$ matrices in spin space.
 
In PM phase $P({\bf m})={\rm const}$,  
the averaging in Eq. (\ref{cpa}) can be 
performed explicitly,   
$\hat{\Sigma}=\Sigma\hat 1$, $\hat{G}=g\hat 1=g_0(E-\Sigma)\hat 1$, 
where $\hat 1$ is a unity matrix, and we obtain
\begin{eqnarray}
g(E)= \frac{1}{2}\sum_{(\pm)} \frac{1}
{g^{-1}(E)+\Sigma(E)
\pm J}.
\label{cpa2}
\end{eqnarray}

To first approximation of the DMFA, leading to Eq. (\ref{cpa}), 
has a simple physical meaning. We reduce the problem of electron scattering due
to many spins, each with the scattering potential 
$-J {\bf m}\cdot {\bf\sigma}$, 
to a problem of a scattering due to a single spin with the effective 
scattering potential
$-J {\bf m}\cdot {\bf \sigma}-\hat{\Sigma}$, 
embedded in an effective medium, 
described by the Hamiltonian $t_{\bf k}+\hat{\Sigma}$,
and, hence, by locator $\hat{G}_{\rm loc}$. 

The same MF approach leads to the second  
DMFA approximation - the approximation for the one-site
probability
$P({\bf m})$, which allows to perform averaging  in FM phase.  
Consider again a single spin with the effective scattering potential  
in an effective medium.
The change in the number of states of the electron gas due to 
such spin is \cite{ziman,hewson} 
\begin{eqnarray}
\label{probability4}
\Delta N(E,{\bf m})= -\frac{1}{\pi}{\rm Im}
\ln {\rm det}\left[1+\left(J{\bf m}\hat{\bf \sigma}
+\hat{\Sigma}_+\right)\hat{G}_{\rm loc\;+}\right],
\end{eqnarray}
where $Y_+\equiv Y(E+i0)$. 
So the change in thermodynamic potential is  \cite{doniach,chat,ak2}
\begin{eqnarray}
\beta\Delta\Omega({\bf m})=\int f(E)\Delta N(E,{\bf m})dE,
\label{probability}
\end{eqnarray}
where  $f(E)$ is the Fermi function, 
the chemical potential is found from the equation
\begin{equation}
n=-\frac{2}{\pi}\int_{-\infty}^{\infty}f(E){\rm Im}\;g_+ \; dE,
\end{equation}
and $n$ is the number of electrons per site.
The result for the one-site probability reads:
\begin{eqnarray}
\label{prob2}
P({\bf m})\propto \exp\left[-\beta\Delta\Omega({\bf m})\right].
\end{eqnarray}

Eqs. (\ref{cpa}) and (\ref{prob2}) are the system of two non-linear (integral)
equations for $\hat{\Sigma}(E)$ and $P({\bf m})$, which one should  solve 
to find thermodynamic properties of the model.
However, in linear  approximation with respect to macroscopic magnetization 
${\bf M}$,
we can reduce this complicated system
to a traditional MF equation for 
${\bf M}$ \cite{aus02}:
\begin{eqnarray}
P({\bf m})\propto  \exp\left( -3\beta T_c{\bf M}\cdot{\bf m}\right).
\label{probability2}
\end{eqnarray}
The parameter $T_c$ is formally introduced as a coefficient in the 
linear term of the expansion of $\Delta \Omega({\bf m})$ with respect to
${\bf M}$ (the reason for the notation we have chosen and for the numerical
coefficient 3 will be clear immediately);  it is determined by the 
properties of the system in paramagnetic phase.
Non-trivial solution of the MF equation 
\begin{equation}
{\bf M}=\left\langle{\bf m}\right\rangle
\end{equation}
can exist only for $T<T_{c}$, hence $T_c$ the Curie
temperature.

\section{$T_c$ for semi-circular DOS}

For simplicity consider the semi-circular (SC) bare density of states (DOS)
$N_0(\varepsilon)$,  the bandwidth being $2W$.
Then
\begin{eqnarray}
\label{gint}
g_0(E)=\int \frac{N_0(\varepsilon)d \varepsilon }{E - \varepsilon}
=\frac{2}{W}\left[\frac{E}{W}-
\sqrt{\left(\frac{E}{W}\right)^{2}-1}\right].
\end{eqnarray}
For this case
\begin{equation}
\label{sigma}
\hat{\Sigma}(E)=E-2w\hat{G}_{\rm loc}-\hat{G}_{\rm loc}^{-1},
\end{equation}
where $w= W^2/8$,
and Eq.  (\ref{cpa}) and  (\ref{cpa2}) take respectively the form
\begin{eqnarray}
\label{sigma2}
\hat{G}_{\rm loc}(E)=\left\langle\frac{1}{E-2w\hat{G}_{\rm loc}(E)
+J{\bf m}\cdot\hat{\bf \sigma}}\right\rangle\\
\label{rq2}
g(E)=\frac{1}{2}\sum_{(\pm)}\frac{1}{E-2wg(E)\pm J}.
\end{eqnarray}
Expanding Eq. (\ref{sigma2}) and then Eq. (\ref{probability4}) with respect to
${\bf M}$, after straightforward algebra,  for the $T_c$  we obtain
\begin{eqnarray}
\label{Theta}
T_{\rm c}=\frac{4J^2w}{3\pi}\int_{-\infty}^{\infty}f(E)\nonumber\\
\mbox{Im}\left[\frac{g_+^2}
{(E_+-2wg_+)(E_+-4wg_+) -\frac{4J^2w}{3}g_+^2}\right]dE.	
\end{eqnarray}
 Formally speaking because the
integral in Eq. (\ref{Theta}) 
contains Fermi function critical temperature enters also into the r.h.s. of the equation.  
But in all cases $T_c$ turns out to be much less the chemical potential, so we
can consider electron gas as degenerate, and Eq. (\ref{Theta}) is an explicit
formula for calculation of $T_c$.  

Eq. (\ref{Theta}) is the main result of our paper. The analysis of this equation 
let us start from the limiting case $J=\infty$. In this case integral can be
calculated explicitly and we obtain \cite{ak2}
\begin{eqnarray}
\label{tcw}
T_c =\frac{W\sqrt{2}}{4\pi}\left[ \sqrt{1-y^{2}}-\frac{1}{\sqrt{3}}\tan^{-1} 
\sqrt{3(1-y^{2})}\right],
\end{eqnarray}
where $y$ is an implicit function of concentration, given by equation 
$n=\frac{1}{2}-\frac{1}{\pi}\left(\sin^{-1}y+y\sqrt{1-y^2}\right)$.
The result coincides with that known previously \cite{furukawa2}.

For arbitrary exchange the integral can be calculated only numerically, but
before we present the results of calculations we should state that
in part of the $J/W-n$ plane  Eq. (\ref{Theta}) 
gives $T_c<0$.
In fact, like in any MF theory of a second order phase transition, 
in our calculation of the critical temperature we
started from a high temperature paramagnetic phase and decreasing 
the temperature were looking
for an instability of a model with respect to  appearance of small 
spontaneous magnetic moment that is for the appearance
of a nontrivial solution  of the MF equation. Negative $T_c$ in some
 part of the
$J/W-n$ plane means, that  at
any temperature, including
$T=0$, the paramagnetic phase here 
is  stable with respect to the appearance
of small spontaneous magnetic moment even at $T=0$, and excludes ferromagnetism
in that region of the plane.  The part of the critical surface corresponding to
the region  of the parameters plane where $T_c\geq 0$ is presented on FIG. 1.
\begin{figure}
\epsfxsize=3truein
\centerline{\epsffile{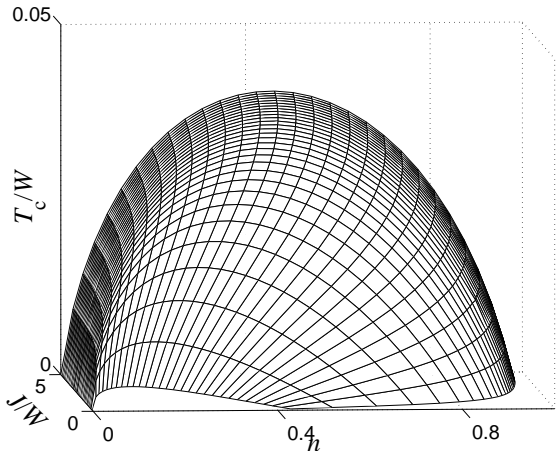}}
\label{FIG.1}
\caption{$T_c$ as a function of
relative strength of the Hund exchange $J/W$ and electron concentration $n$.} 
\end{figure}

\section{Discussion}

Let us finally discuss, whether the PM-FM transition observed with decreasing
temperature can be preceded by 
the transition from the the PM phase to some  magnetic phase other 
than FM (say, antiferromagnetic)?  
We would like to 
present some heuristic arguments that this is not the case.

First, consider the case of weak exchange $J\ll W$. In this case Eq.
(\ref{Theta}) takes the form
\begin{eqnarray}
\label{rkky2}
T_{\rm c} = \frac{2J^2}{3}\int_{-\infty}^{\infty}f(E)
\left\{\frac{dN_0(E)}{dE}-\frac{1}{\pi }\mbox{Im}\;g_0(E_+)^2\right\}dE.
\end{eqnarray}
In fact, this equation  is the MF approximation 
\cite{aus02} for the RKKY Hamiltonian \cite{vf}, to which the original
Hamiltonian
(\ref{HamDXM}) can be reduced to in the case $J\ll W$. So  Eq. (\ref{rkky2}) does not
involve either the coherent potential approximation (Eq. (\ref{cpa})), 
or the SC  density of states. 

Anyhow,  for the SC density of states we use, Eq. (\ref{rkky2}) 
gives ferromagnetic ground state for $n<0.4$, which qualitatively 
agrees with the result of 
numerical calculations,
giving  FM ground state for 
$n<0.25$ for the three principal cubic lattices \cite{mattis}. 

To formulate the second argument,
let us compare the energies of the   PM state 
\begin{equation}
E_p=-\frac{2}{\pi}\int_{-\infty}^{E_F^{(p)}}E\;{\rm Im}\; g_+\; dE;
\end{equation}
and of the saturated FM state 
\begin{eqnarray}
E_f=\int_{-\infty}^{E_F^{(f)}}E\;\sum_{(\pm)}N_{0}(E\pm J)\;dE,
\end{eqnarray}
where $E_F$ is the appropriate Fermi energy.
On fig. 2 we plotted simultaneously the curve given by  by equation
$E_p=E_f$ 
and by equation
$T_c=0$,
where $T_c$ is obtained from Eq. (\ref{Theta}).
\begin{figure}
\epsfxsize=2.2truein
\centerline{\epsffile{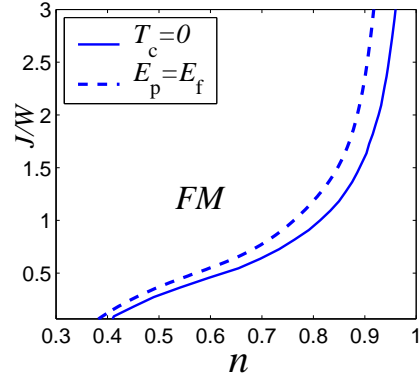}}
\caption{The FM region boundary 
in the coordinates of
relative strength of the Hund exchange $J/W$ and electron concentration $n$.} 
\end{figure}
The close vicinity of the abovementioned curves supports the belief that 
the curve $T_c=0$ is the  
quantum critical line (see \cite{sachdev} and references therein),
which bounds {\it ferromagnetic} phase.
Also, the boundary of ferromagnetic phase on FIG. 2 agrees with those obtained on the basis 
of numerical calculations \cite{dag} and
from qualitative reasoning \cite{chat}.

The destruction of ferromagnetic ground state occurs because finite double
exchange between the itinerant electrons and core spins, unlike an 
infinite one, by itself generates effective antiferromagnetic
exchange between the core spins (which was absent in original Hamiltonian). 
Due to the fact that our main
result (equation for the transition temperature) indicates it's own 
limits of validity, we can find the
boundaries of ferromagnetic phase without analyzing what phases 
are beyond the boundaries.

Finally, we would like to mention, that in the DMFA,
as one can easily see from Eq. (\ref{cpa}), 
density of electron
states in paramagnetic phase does not depend upon  electron concentration. In
this case,
the derivative of chemical potential with respect to number of electrons
 is just the inverse density of states at the
Fermi level (for the degenerate electron gas), and  
 is always positive. Hence,  there is no phase separation in
paramagnetic phase. 

In conclusion, we explicitly formulated the Dynamical Mean Field Approximation
equations
for the double exchange model with classical spins
 for arbitrary relation between Hund exchange and
the electron bandwidth. Near paramagnetic-ferromagnetic transition critical
point, these equations were reduced to a MF equation, describing a
single spin in an effective field, proportional to the macroscopic
magnetization. The effective exchange interaction, 
entering into the MF equation was
found for the semicircular electron density of states. We thus calculated the
transition temperature $T_c$ as a function of Hund exchange interaction and
electron density in the whole parameters plane. The results obtained also allow
to plot the boundaries of the ferromagnetic region on the model phase diagram.

This research was supported by the Israeli Science Foundation administered
by the Israel Academy of Sciences and Humanities and BSF.

\end{multicols}
\end{document}